\documentclass[reprint,
 amsmath,amssymb,
 aps,
]{revtex4-2}
\usepackage{svg}
\usepackage{comment}

\usepackage{graphicx}
\usepackage{dcolumn}
\usepackage{bm}
\usepackage{physics}
\usepackage{textcomp}
\usepackage{siunitx}
\svgsetup{inkscapeexe = "C:/Program Files/Inkscape/bin/inkscape.exe"}
\usepackage{float}
\begin{document}


\title{Mie Voids as broadband directional light sources}

\author{Benjamin Reichel\textsuperscript{1}}
\email{benjamin.reichel@pi4.uni-stuttgart.de}
\author{Adrià Canós Valero\textsuperscript{2}}%
\author{Mario Hentschel\textsuperscript{1}}
\author{Harald Giessen\textsuperscript{1}}
\author{Thomas Weiss\textsuperscript{2}}

\affiliation{\textsuperscript{1}4\textsuperscript{th} Physics Institute and SCoPE, University of Stuttgart, Pfaffenwaldring 57, 70569, Stuttgart, Germany}
\affiliation{\textsuperscript{2}Institute of Physics, University of Graz, and NAWI Graz, Universitätsplatz 5, Graz 8010, Austria}

\date{\today}

\begin{abstract}
The Kerker effect arises from the interference between electric and magnetic multipoles, enabling directional light scattering in nanophotonics. However, conventional dielectric and plasmonic nanoparticles can only act as Kerker sources in narrow spectral regions, limiting their applicability. Here, we show that the recently discovered Mie voids overcome this limitation by supporting a broadband generalized Kerker effect spanning the whole visible range. We investigate the optical response of Mie voids under both plane-wave and dipolar excitation. For plane waves, the voids preferentially scatter light in the forward direction. Under dipolar excitation,
the resulting radiation emission towards the void and beyond is suppressed due to destructive interference between the dipole field with the directional scattered field of the void. These findings identify Mie voids as versatile broadband directional sources, opening pathways for antenna design and energy harvesting at the nanoscale.

\begin{description}
\item[DOI] ---
\item[Keywords]
Mie Voids, Mie Theory, Metamaterials, Dielectric Nanostructures, Purcell Factor
\end{description}
\end{abstract}

\maketitle



\section{Introduction}
When a localized current distribution radiates through a balanced combination of electric and magnetic dipole responses, the emitted fields  can acquire a directional character. In electromagnetic scattering, the conditions for this behavior are known as the Kerker conditions, introduced by Kerker in 1983 in the context of spheres with equal permittivity and permeability~\cite{Kerker:83}. The first Kerker condition corresponds to in-phase electric and magnetic dipoles, resulting in the suppression of backward scattering, while the second condition corresponds to out-of-phase dipoles, suppressing forward scattering. In nanophotonics, the concept remained largely theoretical due to a lack of magnetic materials~\cite{liu2018generalized}, until the advent of metamaterials and the discovery of optically induced magnetic resonances in high-index dielectrics enabled the first experimental realizations~\cite{PhysRevApplied.21.024028,Verre2019,Green:20,798002,zhao:hal-00471883,Kuznetsov2012,  Evlyukhin2012,Liu_2014,Geffrin2012,Fu2013,Person2013,9364857}. These developments have led to a broad range of applications of the Kerker effect for tailoring light directionality at the nanoscale ~\cite{Shakirova_2021,Geffrin2012,Fu2013,zhao:hal-00471883,Liu:14,doi:10.1126/science.aag2472,Liu_2017,Staude2017,Kivshar:17,YANG20171,Kruk2017,Ding_2018,Neugebauer2016,10.1117/12.2593441}. While the Kerker conditions were initially developed for the simplified case of a single sphere under plane wave 
illumination in free space, today, the inclusion of higher-order multipolar excitations or illumination types give rise to generalized Kerker conditions, where directionality emerges from far-field interference among several multipoles, beyond the dipole approximation~\cite{Shakirova_2021,Alaee:15,Neugebauer2016,Hancu2014,Li2016}.

Crucially, all realistic nanostructures that support the Kerker conditions do so through resonances~\cite{Shakirova_2021,Staude2013,Luk’yanchuk2015}: the exact cancellation or enhancement of a particular scattering direction typically occurs only in a narrow spectral window where the relevant multipolar resonances spectrally coincide. In practice, this means conventional Kerker scattering is limited to very narrow frequency bands, restricting its usefulness for broadband or multiwavelength applications.

\begin{figure}
\includesvg{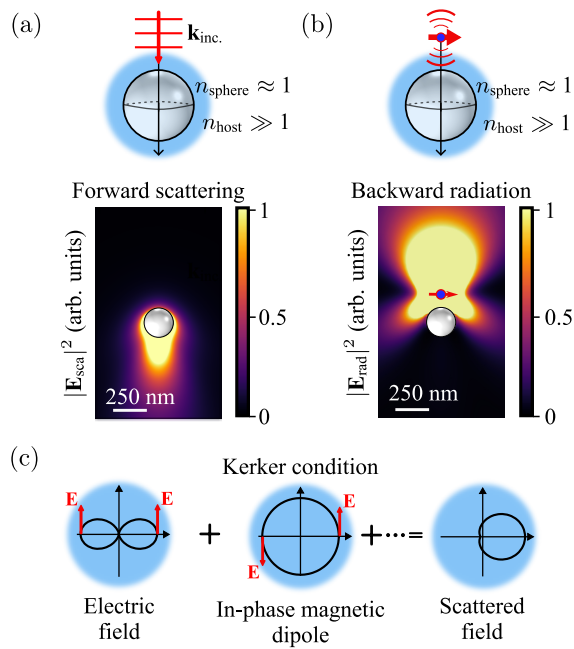}
\caption{Illustration of the scattering and radiation behavior of spherical Mie voids under different kinds of illumination: (a) plane wave excitation. (b) Directional Purcell enhancement effect in the case of a dipole excitation. We denote the emitted radiation towards the dipole and beyond as `foward radiation'. The emitted forward radiation, which is the linear superposition of the scattered and dipole fields, is suppressed, whereas the radiation to the opposite or backward direction remains. (c) Generalized forward Kerker effect of spherical Mie voids in the case of plane waves. The far-field emission is predominantly in the forward direction.}
\label{fig: Headfigure}
\end{figure}

In this work, we introduce a new approach to overcome these limitations. Instead of relying on high-index resonator structures, we consider the recently discovered \emph{Mie voids}, which are air cavities embedded in a high-index material~\cite{Hentschel2023, PhysRevLett.130.243802}. These voids possess resonant modes localized in the low-index region and are the complementary counterparts of solid dielectric spheres~\cite{Hentschel2023}. Owing to this confinement in air, their modes inherit antenna-like features while avoiding the loss and dispersion of the dielectric host. However, the theoretical understanding of Mie voids remains limited. Recent studies have shown that their mode spectra can be related to those of solid resonators via a quasi-Babinet principle~\cite{Hentschel2023,PhysRevResearch.7.013136}. Yet, their scattering response under realistic excitation conditions is still largely unexplored. Here, we demonstrate that Mie voids act as broadband, unidirectional scattering sources governed by generalized Kerker conditions. We find that, unlike solid spheres, the scattering of Mie voids is dominated by the coherent interference of multiple spectrally overlapping multipoles. Remarkably, as shown in Fig.~\ref{fig: Headfigure}(a,b), this directionality is robust under both far-field and near-field excitation (plane-wave and dipole illumination).

Under plane-wave illumination, this results in a strongly suppressed backscattering and dominant forward scattering. Under local dipole excitation, two distinct regimes emerge. When the emitter is placed inside the void, coupling to the resonant modes in air yields a higher spontaneous emission rate of the dipole compared to its emission in free space, despite the emitter being located in air~\cite{Purcell1946,Novotny_Hecht_2012}. This emission rate modification compared to emission in a homogeneous background is quantified by the Purcell-factor~\cite{PhysRevE.62.4251,Purcell1946,Novotny_Hecht_2012}, which can reach values up to 5 for Mie voids, see Fig.~\ref{fig: Purcell enhancement}. When the emitter is outside the void, the resulting radiation emission towards the void and beyond is suppressed due to destructive interference between the dipole field with the directional scattered field of the void.  The radiation towards the void and beyond defines the forward direction in the case of dipole illumination. These effects, rooted in broadband multipolar interference, highlight the potential of Mie voids as highly directive nanoantennas for both forward and backward emission control, opening opportunities for coupling light to deeply embedded emitters and for enhancing optical functionalities in high-index nanophotonic platforms.

\section{\label{sec: Multipole-Decomposition} Theoretical background}

We mainly consider scattering by a spherical Mie void of radius $R$  and refractive index $n_{\text{void}}\approx 1$, embedded in a host medium with refractive index $n_{\text{host}} \gg 1$. The system is excited by either an incident plane wave or a point-dipole source, as sketched in Fig.~\ref{fig: Headfigure}(a) and Fig.~\ref{fig: Headfigure}(b). 
Following standard Lorenz-Mie theory~\cite{jackson1998classical,mie1908,bohren1983chap3and4,Hohenester2020}, the scattering response of the void can be obtained by calculating the contributions of the electric and magnetic multipolar fields, given by the Mie coefficients $a_l$ and $b_l$, respectively, see Appendix \ref{Ap: MultipoleDecomposition}. 
The quantum number $l$ determines 
the order of the multipole, where $l=1$ are dipoles, $l=2$ quadrupoles, and so on. 

The Kerker effects arise from the opposite spatial symmetries of electric and magnetic multipoles of the same order~\cite{liu2018generalized}. The electric field of the electric dipole in the forward and backward directions (with respect to the incident illumination) are in phase. In contrast, the electric field of the magnetic dipole always has an opposite sign in the forward and backward directions.

As a result, if the fields of the electric and magnetic dipole are in-phase, the total scattered field will be enhanced in the forward direction, while backward scattering will be suppressed, as shown in Fig.~\ref{fig: Headfigure}(c).

While the Kerker effect is mostly known for dipolar fields, the same symmetry argument holds for all multipolar orders~\cite{dezert:tel-02869591,liu2018generalized}. If higher-order multipole fields are involved, this effect is known as the generalized Kerker effect~\cite{Alaee:15,liu2018generalized}. In the remainder of this work, we will refer to the generalized Kerker effect simply as Kerker effect.

\begin{figure}
	\includesvg{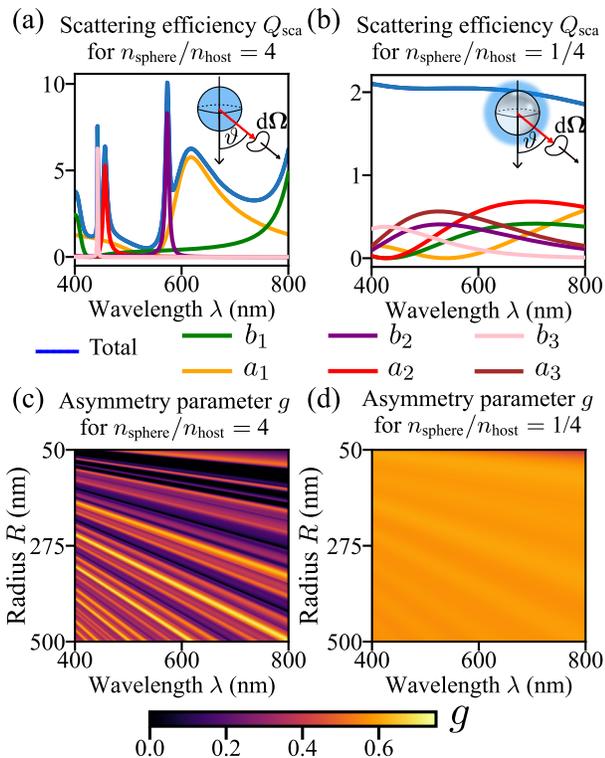} 
	\caption{ (a)  and (b): The first six multipole contributions to the scattering efficiency $Q_{\text{sca}}$ for a Mie void with $\SI{100}{\nano\meter}$ radius as a function of the vacuum wavelength $\lambda $. The scattering angle $\vartheta$ is defined in the upper right corner of each panel. (c) and (d): The asymmetry parameter $g$  as a function of $R$ and $\lambda$. Panels (a) and (c), respectively, display $Q_{\text{sca}}$ and $g$ for a `conventional' Mie resonator case $n_{\text{sphere}} >n_{\text{host}}$, whereas  the results in panels (b) and (d) have been derived  for Mie voids with $n_{\text{sphere}} <n_{\text{host}}$ .}
	\label{fig: scattering cross section}
\end{figure}

In general, the scattering directionality can be quantified  by the so‑called asymmetry scattering parameter 
\begin{equation}
	g = \langle \cos(\vartheta) \rangle = \frac{1}{I_{\text{inc}}} \int_{S_2} \cos(\vartheta)\, \left(\vb{S}_{\text{sca}}\cdot \vb{n}\right)r^2 \; \dd \Omega,
\end{equation}
where $\vartheta$ is the scattering angle of the outgoing field, which is depicted in Fig.~\ref{fig: scattering cross section}(a,b),  and $I_{\text{inc}}$ is the incoming irradiance~\citep{bohren1983chap3and4}. 
 This weighted cosine function is between -1 corresponding to backward scattering $(\vartheta = \pi)$ and 1 corresponding to forward scattering $(\vartheta=0)$. 
 With the help of the multipole expansion series, we can write $g$ again as a sum of $a_{l}$ and $b_{l}$~\cite{bohren1983chap3and4}, which reads 
 \begin{align}
 \begin{split}
g &= \frac{4}{(kR)^2\,Q_{\mathrm{sca}}} \sum_{l=1}^{\infty} \frac{2l+1}{l(l+1)} \Re(a_{l}b_{l}^*) \\ 
		&+\frac{l(l+2)}{l+1} \Re \left(a_{l}a_{l+1}^* + b_{l}b_{l+1}^*\right), 
		 \label{eq:g}
 \end{split}
 \end{align}where we have also introduced the particle scattering efficiency~\cite{bohren1983chap3and4,Gouesbet2023} as a function of relative refractive index  $\eta = n_{\text{sphere}}/n_{\text{host}}$ and the size parameter $kR$
 \begin{align}
 	\begin{split}
 	Q_{\text{sca}} &=  \frac{2}{ (kR)^2} \sum_{l}(2l+1)\left( |a_{l}(\eta,kR)|^2 + |b_{l}(\eta,kR)|^2 
 	\right),
 	\end{split}
 	\label{eq: Q_sca}
 \end{align}
   we can safely conclude that for $g> 0.5,$ higher-order multipole contributions are responsible for the additional forward scattering. 
\newpage
\section{\label{sec: Kerker effect in Mie Voids} Kerker effect in spherical Mie voids}

We start our investigation by comparing $Q_{\text{sca}}$ of conventional Mie spheres and spherical Mie voids (with $R=\SI{100}{\nano \meter}$). 
  For Mie spheres, the relative index is  
$\eta>1 $, whereas for  Mie voids, $\eta<1$. For illustration purposes, 
we choose  $\eta=4$ and $\eta=1/4$ for the Mie sphere and Mie void, which is a typical ratio between a semiconductor and vacuum in the visible and infrared spectral region.

Figure~\ref{fig: scattering cross section}(a) displays the numerical results of $Q_{\text{sca}}$ for optical wavelengths $\lambda$ in the range of $\SIrange[range-phrase=\text{ to },range-units=single]{400}{800}{\nano\meter}$ for Mie spheres. We can see sharp narrow-banded peaks for $Q_{\text{sca}}$ where the individual $a_l$- and $b_l$-multipole contributions, i.e., the summands in Eq.~\eqref{eq: Q_sca}, only have weak spectral overlap. Therefore interference between multipole fields remains limited. A broadband generalized Kerker condition is not possible due to a lack of spectrally overlapping multipole fields for a $\SI{100}{\nano\meter}$ Mie sphere and optical wavelengths. However, it is actually possible to obtain spectrally overlapping $a_l$- and $b_l$-multipole contributions, if the size parameter $kR$ is increased~\cite{vandehulst1981,bohren1983chap3and4,jackson1998classical}, i.e, by either increasing $R$ or decreasing $\lambda$.   

In Fig.~\ref{fig: scattering cross section}(b), $Q_{\text{sca}}$ is almost constant around two, which is due to the lower quality factor of void modes in that spectral range~\cite{Hentschel2023,PhysRevResearch.7.013136}, so that the $a_l$- and $b_l$-multipole contributions are broadband for  Mie voids ~\cite{PhysRevResearch.7.013136}.
As a result, they spectrally overlap across the optical wavelengths. 
Moreover, all non-zero high-order $a_l$- and $b_l$-multipole contributions are
roughly of the same relative magnitude.  
For the sake of simplicity,
we only  plot them up to the third order multipole terms
$a_3$ and $b_3$, respectively, although higher-order multipoles are
also involved in the scattering.
As a result, a Mie void with a $\SI{100}{\nano\meter}$ radius easily exhibits the aforementioned Kerker
effect~\cite{PhysRevResearch.7.013136} in the optical range owing to the multitude of spectrally overlapping multipole fields, unlike those of solid spheres.

Switching to the discussion of the asymmetry parameter $g$, Fig.~\ref{fig: scattering cross section}(c)
displays $g$ for a high indexed sphere for different radii $R$ and wavelengths $\lambda$. We can see that $g$  also has multiple sharp narrow-banded peaks, which appear for a broad range of  $R$ around $\SIrange[range-phrase=\text{ to },range-units=single]{50}{500}{\nano\meter}$.    
This illustrative example
for high-indexed resonators just reaffirms the fact
that the Kerker condition is only met for few possible ($R,\lambda$)  values, where $g$ reaches values greater than 0.5. 

In strong contrast, Fig~\ref{fig: scattering cross section}(d) displays $g$ for a Mie void with a fixed $\eta=1/4$.
Instead of a strongly varying $g$, we have a robust, almost flat $g \geq 0.5,$ which indicates a generalized Kerker effect since this is only possible with the help of higher-order multipole excitations, as discussed in Section~\ref{sec: Multipole-Decomposition}. Physically, Eq.~\eqref{eq:g} shows that the contributing factors to the scattering arise from the coupling between different $a_{l}$ and $b_{l}$.  The first term in Eq.\eqref{eq:g} describes coupling between magnetic and electric multipoles of the same order. The second one describes coupling between two adjacent multipole orders of the same type $a_{l}$ and $a_{l+1}$. For situations where $a_l \approx b_l$, the multipole fields will destructively interfere in the backward direction and constructively in the forward direction leading to a Kerker-like response, and we obtain a positive $g$. Since for Mie Voids, there are multiple spectrally overlapping multipole fields, the robust broadband regions with $g\gg 0$ are not surprising. For the analysis of more experimentally relevant systems, we briefly discuss the simulation of a conical Mie void in a zero absorption silicon substrate in the subsequent section. 

Even though we restricted our discussion to specific values of $R$ and $\lambda$ with fixed $\eta$, the same effect can be observed over a wider parameter range. This is demonstrated in Appendix~\ref{Ap: MultipoleDecomposition} by computing  $g$ as a function of $kR$ and $\eta$. In addition, we also show in Fig.~\ref{fig: c_i(m,kR)} that Mie voids exhibit resonant responses in this regime which are related to those of Mie resonators by the quasi-Babinet principle. Surprisingly, despite these resonances, voids feature $g>0.5$ in the whole parameter range studied in Fig.~\ref{fig: G-Paramer(m,kR)}.


\section{Numerical Simulation of non-spherical Mie voids}
The challenge with voids embedded in a homogeneous host medium is that they are difficult to access if the host medium is absorbing. Therefore, we switch to an experimentally more feasible system, namely conical Mie voids with top and bottom radii $r_{\text{top}/\text{bot}}$ and height $h$ embedded in a  substrate medium, which we illuminate by circularly polarized plane waves of wavelength $\lambda$. The medium used in experiments is often silicon or gallium arsenide \cite{Hentschel2023,Arslan_2025}. In order to be close to possible experiments, we assume a refractive index of $n_{\text{host}}=4$, which corresponds to a high-index dielectric for optical wavelengths. The superstrate medium on top is vacuum.
Whereas our analysis of spherical Mie voids provided an analytical solution via the multipole expansion series, closed form solutions for $a_{l}$ and $b_{l}$ for cone-shaped voids are currently unknown. Therefore, we obtain the scattering response of the system using the finite-element solver COMSOL Multiphysics\textsuperscript{\textcopyright}. 

\begin{figure}
	\includesvg{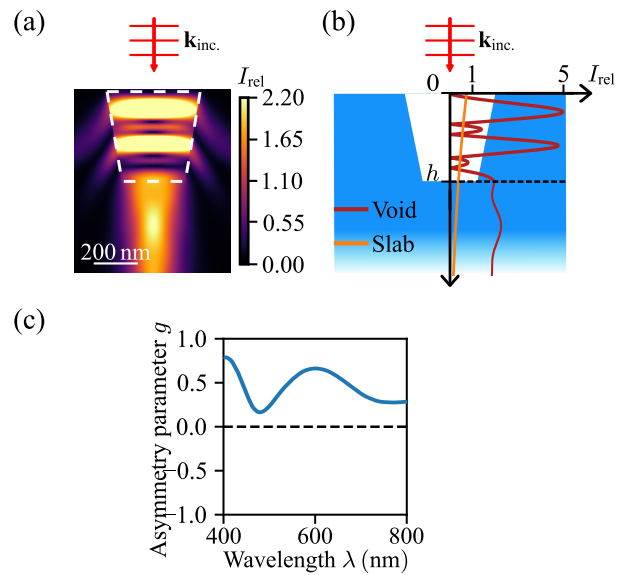}
	\caption{(a)  Numerically calculated intensity distribution $I_{\text{sca}}$ of the scattered electric field around an isolated conical void for an incident plane wave coming from the top at a wavelength of $\lambda=\SI{600}{\nano \meter}$.
		The local field intensity is normalized to $I_{\text{inc}}$ as the intensity of the incident field. The Mie void with upper radius $r=\SI{200}{\nano \meter}$, lower radius $r=\SI{150}{\nano \meter}$, and height $h=\SI{410}{\nano \meter}$ is embedded n a substrate with $n_{\text{host}} = 4$. (b) Penetrated relative intensity $I_{\text{sca}}/I_{\text{inc}}$ along the symmetry
		axis in the center of the void compared to the penetrated relative intensity in the absence of the void.
		(c)  Asymmetry parameter for the same void as a function of $\lambda$.  }
	\label{fig: conical voids2}
\end{figure}

The scattered intensity distribution $I_{\text{rel}}=|\vb{S}_{\text{sca}}|/I_{\text{inc}}$ at a wavelength of 600~nm is depicted in Fig.~\ref{fig: conical voids2}(a). We observe that, as in the spherical void case, there is a forward scattering behavior even for non-spherical Mie voids.  

For further evidence, we depict $g$ for a fixed conical void as a function of the incoming wavelength $\lambda$ in panel~(b). Beginning from the mid-optical wave spectrum and near $\SI{400}{\nano\meter}$, we clearly see forward scattering behavior that reaches values close to $g\approx 0.5$. However, in contrast to the spherical case, for wavelengths near $\SI{500}{\nano\meter}$ and high optical wavelengths, the forward scattering light reflection increases and $g$ goes below 0.5, while always being greater than 0.  
Qualitatively, this behavior is similar to the spherical Mie void case to first order, where we also observe bands in which the $g$ parameter reaches values up to 0.6. As a result, the open geometry inherits its directional properties from the Kerker effect that takes place in the fully embedded voids. 
Additionally, we compare in Fig.~\ref{fig: conical voids2}(b) the scattering behavior of conical Mie voids in a substrate (blue line) to the scattering of a dielectric half-space without a Mie void (orange line). Here $I_{\text{rel}}$ is plotted along the penetration depth as a function of the symmetry axis of the void, which is indicated in Fig.~\ref{fig: conical voids2}(b) by the black arrow pointing into the substrate. We plot the scattered intensity along the symmetry axis. The scattered intensity for the void is extracted from the simulation, while for the dielectric half-space, it is calculated analytically~\citep{bohren1983chap3and4}.
We can see that inside the substrate, i.e., for penetration depths $z > h = \SI{500}{\nano\meter}$, the scattered intensity for a void outperforms the scattering intensity without a void. Therefore, even though the more realistic voids are no longer spherical nor are they fully embedded inside an homogeneous host medium, they still display strong forward scattering. As a result, their scattering response paves the way for possible applications of Mie voids as highly directive nanoantennas radiating into a substrate.
\section{Directional Purcell effect in Mie voids}
In this section, we investigate the potential of Mie voids to control the radiation properties of weakly coupled quantum emitters. The latter are modeled as electric dipoles $\vb{p}$ placed inside or in the vicinity of the void, as sketched in the insets of Fig.~\ref{fig: Purcell enhancement}(a) and Fig.~\ref{fig: Purcell enhancement}(b). As a figure of merit, we consider the Purcell factor $F$ to quantify the enhancement of the spontaneous emission rate by the quantum emitter at a certain spatial position~\cite{Purcell1946}. Without loss of generality, we only need to calculate $F$ for an emitter whose dipole moment is oriented parallel ($F_{\parallel}$) or perpendicular ($F_{\perp}$) to the surface of a Mie void. This can be justified by the fact that for axially-symmetric structures, the Purcell factor for an arbitrary $\vb{p}$ is a linear combination of $F_{\parallel}$ and $F_{\perp}$ according to~\cite{Hohenester2020}
\begin{equation}
	F(\vb{p}) = \sin^2(\vartheta) F_{\perp} + \cos^2(\vartheta) F_{\parallel},
\end{equation}
where $\vartheta$ is the angle between the unit normal of the sphere and the emitter dipole moment. We provide the explicit expressions for $F_{\parallel}$ and $F_{\perp}$ inside and outside the void in Appendix \ref{Ap: Purcell-Factor Calculation}.

\begin{figure}
	\includesvg{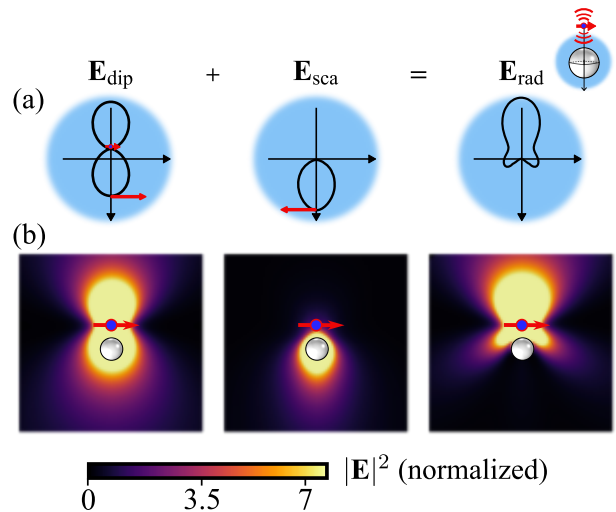}
	\caption{(a) Illustration of the directional Purcell effect in the case of a local dipole illumination. The scattered field of the Mie void destructively interferes with the dipole field, leading to backward radiation. (b) Spatial distribution of the decomposition of the radiated electric field intensities $|\vb{E}_i|^2$ for a $\SI{100}{\nano\meter}$ void in the presence of a dipole with wavelength $\lambda = \SI{600}{\nano\meter}$, which is placed $\SI{160}{\nano\meter}$ away from the void center. The index $i$ indicates the dipolar, the scattered, or the radiated field. If the scattered electric field has a different phase than the dipolar field, this leads to destructive interference. As a result, the radiated field intensity has a minimum in the forward direction.}
	\label{fig: wave picture}
\end{figure}

We first consider the case when the emitter is placed inside the void, as shown in Fig.~\ref{fig: wave picture}. In this scenario, the resonant modes in the air region naturally lead to a significant Purcell enhancement, as discussed in detail in Appendix~\ref{Ap: Purcell-Factor Calculation}.
Interestingly, when  the emitter is placed outside the void, we observe no significant enhancement. Instead, in a similar fashion to the plane-wave scattering problem studied in the previous sections, the radiated field by the emitter acquires a directional character, shown on the rightmost panel of Fig.~\ref{fig: wave picture}(b).    
From a wave perspective, we see this simply as destructive interference between the scattered field of the Mie void $\vb{E}_{\text{sca}}$ and the dipole field $\vb{E}_{\text{inc}}$. Our dipole radiates isotropically in the forward and backward directions, and we know from Sec.~\ref{sec: Kerker effect in Mie Voids} that the Mie void exhibits a generalized forward Kerker effect under plane-wave illumination. Similarly to the latter case, when the emitter is placed outside the void, $\vb{E}_{\text{sca}}$ is predominantly in the forward direction, as shown in the second panel of Fig.~\ref{fig: wave picture}(b). Therefore, the original dipole field by the emitter and the scattered field interfere only in the forward direction, whereas the emitter radiates undisturbed in the backward direction, giving rise to an overall backward radiated field $\vb{E}_{\text{rad}} = \vb{E}_{\text{dip}} + \vb{E}_{\text{sca}}
$, as shown in the rightmost panel of Fig.~\ref{fig: wave picture} panel (b). 

Consequently, placing the emitter outside the void trades away resonant Purcell enhancement for spatial control. The local density of states remains essentially unaltered while the power is strongly emitted into the backward direction. In the future, this spatial control of emission might be exploited to realize directional single-photon sources by judiciously choosing the emitter position and orientation with respect to the Mie void.

\section{summary}

In summary, we have shown that spherical Mie voids in high-index dielectrics exhibit rich scattering behavior that can be understood in terms of broadband multipolar interference. In the case of plane wave illumination, the coherent superposition of multiple spectrally overlapping multipole moments produces a generalized Kerker effect, resulting in strongly suppressed backscattering and highly directional forward radiation. This forward-scattering response persists across the entire optical spectrum and enables significantly enhanced light penetration into the substrate compared to a flat dielectric interface. In contrast, when excited by a local dipole source, Mie voids exhibit two distinct behaviors depending on the emitter's position. For a dipole placed inside the void, coupling to the cavity modes substantially increases the local density of optical states, leading to a Purcell factor enhancement. For a dipole located outside the void, destructive interference between the scattered and dipolar fields suppresses forward emission and directs radiation predominantly backward. Taken together, these findings establish Mie voids as highly directive nanoantennas, capable of converting incident plane waves into forward-directed scattering and enforcing backward emission from nearby emitters. The ability to control emission direction through multipolar interference opens avenues for advanced light–matter interaction schemes, including addressing deep emitters, directional single-photon sources, sensing and enhanced energy harvesting in dielectric nanostructures.

\begin{acknowledgments}
B.R. acknowledges Sergei Gladyshev for providing the
initial simulation codes. This work was supported by Bundesministerium für Bildung und Forschung,
Deutsche Forschungsgemeinschaft, (SPP1839 “Tailored Disorder” and GRK2642 “Towards Graduate Experts in Photonic
Quantum Technologies”), the Ministerium für Wissenschaft,
Forschung und Kunst Baden-Württemberg (RiSC Project
“Mie Voids”, ZAQuant) and Vector Stiftung MINT Innovationen.
B.R. and A.V. contributed equally to this work. M.H.
and H.G. conceived the original idea. B.R. performed the simulations and modeling. B.R.,  A.V. and T.W. derived the theoretical explanation. M.H. and H.G. provided useful insights in the experimental realization. 
All authors participated in the preparation and writing of the
manuscript. A.V. and T.W supervised the work.
\end{acknowledgments}
\appendix
\section{Multipole Expansion and Mie Theory}  
\label{Ap: MultipoleDecomposition}
In this section we provide a discussion of the asymmetry parameter $g$ as a function of the host size parameter $k R$ and the relative refractive index $\eta = n_{\mathrm{sphere}} / n_{\mathrm{host}}$. Here, $n_{\mathrm{host}}$ and $n_{\mathrm{sphere}}$ denote the refractive indices of the host medium and the sphere, $R$ is the sphere radius, and $k = 2 \pi n_{\mathrm{host}} / \lambda$ is the host wavenumber.

Following standard Mie theory~\cite{mie1908} for non magnetic particles, the scattering multipole coefficients $a_l$ and $b_l$, together with the internal coefficients $c_l$ and $d_l$, are given by
\begin{gather}
	a_l =	
	\frac{
		\eta \psi_l(\eta kR) \psi'_l(kR)
		-
		 \psi'_l(\eta kR) \psi_l(kR)
	}{
		\eta \psi_l(\eta kR) \xi'_l(kR)
		-
		 \psi'_l(\eta kR) \xi_l(kR)
	},
	\\[6pt]
	b_l =	
	\frac{
		\eta \psi'_l(\eta kR) \psi_l(kR)
		-
		 \psi_l(\eta kR) \psi'_l(kR)
	}{
		\eta \psi'_l(\eta kR) \xi_l(kR)
		-
		 \psi_l(\eta kR) \xi'_l(kR)
	},\\[6pt]
	c_l =	
	\frac{
		\eta\xi'_l(kR) \psi_l(kR)
		-
		\eta \xi_l(kR) \psi'_l(kR)
	}{
		\eta \psi_l(\eta kR) \xi'_l(kR)
		-
		 \psi'_l(\eta kR) \xi_l(kR)
	},\\[6pt]
	d_l =
	\frac{
		\eta \psi_l(kR) \xi'_l(kR)
		-
		\eta \psi'_l(kR) \xi_l(kR)
	}{
		\eta \psi'_l(\eta kR) \xi_l(kR)
		-
		 \psi_l(\eta kR) \xi'_l(kR)
	},
	\label{eq: generalized MultipoleCoefficients}
\end{gather}
where we have introduced the abbreviations $\psi_l(x)=xj_l(x), \xi_l(x)= x h_l(x)$ for the spherical Riccati-Bessel and Riccati-Hankel functions, with primes indicating derivatives with respect to the argument. For the derivations of the multipole coefficients, see~\cite{bohren1983chap3and4}. With the help of $a_l$ and $b_l$, we can compute $g$ as a function of $kR$ and $\eta$ for spherical particles, see Eq.~\eqref{eq:g} . 

Figure~\ref{fig: G-Paramer(m,kR)} displays the resulting $g$, for size parameters between 1 and 50 as well as the index ratios between 0.1 and 5. As in Fig.~\ref{fig: scattering cross section}(c), in the Mie resonator regime, i.e., $n_{\mathrm{sphere}} > n_{\mathrm{host}}$, $g$ exhibits multiple sharp peaks for high-index resonators. Consequently, the Kerker condition is satisfied only for a few parameter pairs $(k R, \eta)$, where $g$ exceeds about 0.5.

As $n_{\mathrm{sphere}}$ approaches $n_{\mathrm{host}}$, $g$ increases and can approach unity in the limit $\eta \rightarrow 1$. This is expected because for particles with a refractive index close to the surrounding medium, the particle acts as a weak perturbation to the incident wave. As a result, the scattered field remains nearly in phase with the incoming wave. This weak contrast also suppresses higher-order multipole  fields~\cite{jackson1998classical}, leading to predominantly forward scattering due to constructive interference in the forward direction and destructive interference in the backward direction~\cite{olmos2020optimal}.

In the Mie void regime, if the size parameter does not approach 1, $g$ is robust, nearly flat, and typically larger or around $0.5$. This indicates a generalized Kerker effect since this is only possible with the help of higher-order multipole excitations. 

\begin{figure}
	\includesvg{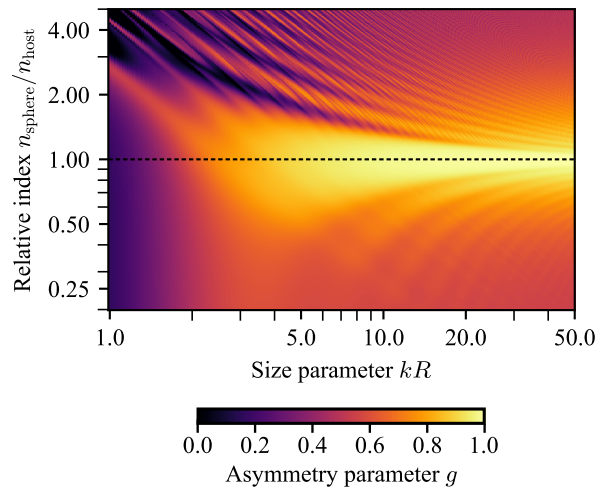}
	\caption{The asymmetry
		parameter $g$ for spherical particles excited by plane waves. $g$ is plotted as a function of the size parameter $kR = 2\pi n_{\text{host}}R/\lambda$ and the relative refractive index $n_{\text{sphere}}/n_{\text{host}}$.The boundary between Mie-void and Mie resonator regimes is displayed by the dashed black line}
	\label{fig: G-Paramer(m,kR)}
\end{figure}
This interpretation is further supported by an analysis of the internal coefficients $c_l$ and $d_l$.
While the scattering multipole coefficients $a_l$ and $b_l$ are crucial for computing 
$Q_{\text{sca}}$ and $g$, the internal coefficients $c_l$ and $d_l$ primarily reveal the the resonant behavior of the particle~\cite{PhysRevResearch.7.013136,bohren1983chap3and4}. 

\begin{figure}
	\includesvg{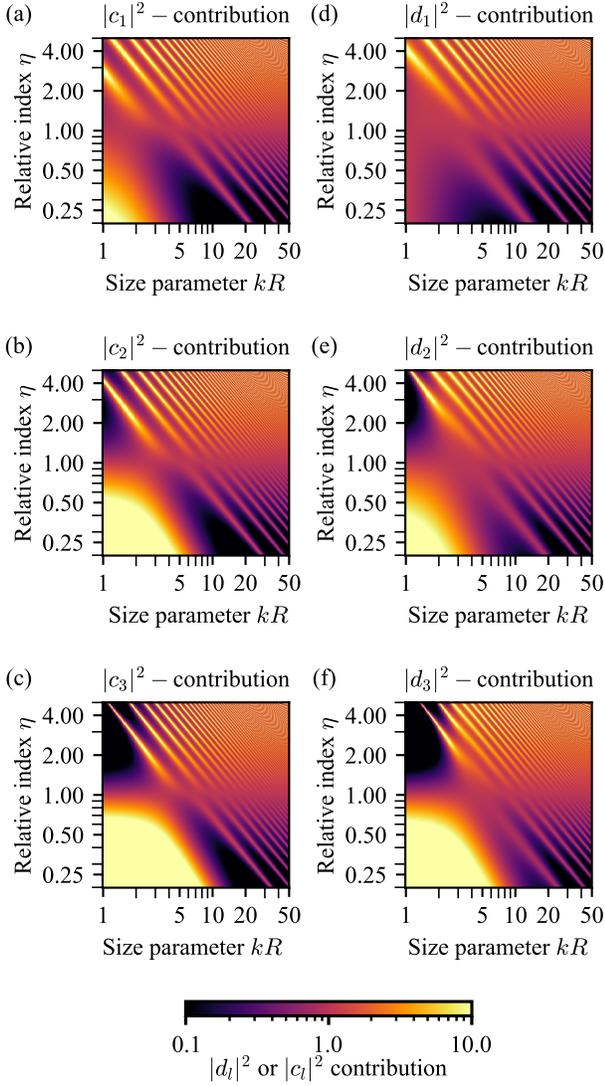}
	\caption{(a - c) $|c_l|^2,l\in \{1,2,3 \}$ as a function of the relative refractive index $\eta= n_{\text{sphere}}/n_{\text{host}} $ and host size parameter $kR = 2\pi n_{\text{host}}R/\lambda$ for spherical particle under plane wave excitation. (d - f) Same as in (a - c), but for $|d_l|^2,l\in \{1,2,3 \}$.}
	\label{fig: c_i(m,kR)}
\end{figure}

Figure~\ref{fig: c_i(m,kR)} displays $|c_l|^2$ and $|d_l|^2$ for $l\in \{1,2,3 \}$ as functions of the size parameter $kR$ and $\eta$ under plane-wave excitation. 

In the Mie resonator regime, i.e., $n_\text{sphere}>n_\text{host}$, we observe sharp resonances, which sharpen for higher $\eta$. This is expected from high-index resonators. 

In the Mie-void regime, the first three internal multipole coefficients also exhibit resonances, showing that the large $g$ values in Fig.~\ref{fig: G-Paramer(m,kR)} are associated with resonances of the Mie voids, which differ from those of Mie resonators. In particular, a resonance peak of $|c_l|^2$ in the Mie resonator regime (as a function of $\eta$ and $k R$) does not generally persist into the void regime; instead, it appears as a $|d_l|^2$ resonance, and conversely.

This behavior can be explained by the quasi-Babinet principle~\cite{PhysRevResearch.7.013136}, which relates the electric and magnetic fields of a structure to those of its complementary counterpart and links their resonant frequencies and quality factors. The principle holds quantitatively for spherical voids or resonators with sizes larger than the wavelength in the host medium, and approximately for sub-wavelength structures.

Finally, in the small-size limit, the internal coefficients scale as $|c_l|^2 \sim (n_{\mathrm{host}} / n_{\mathrm{sphere}})^{2l}$. This explains the distinct small-size behavior of Mie voids and resonators. For Mie voids, this corresponds to the formal limit of the sphere having a vanishing refractive index ($n_{\mathrm{sphere}} \to 0$) or, equivalently, a host index $n_{\mathrm{host}} \to \infty$.

\section{Calculation of the Purcell factor }
\label{Ap: Purcell-Factor Calculation}

\begin{figure}
\includesvg{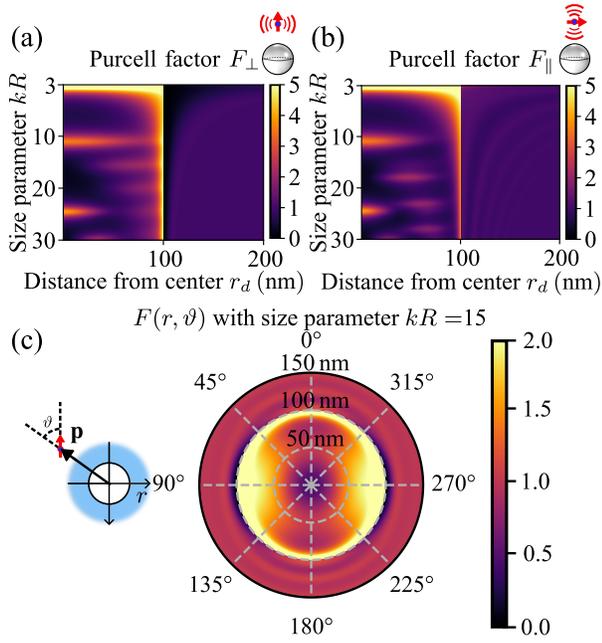}
\caption{(a), (b) Purcell enhancement factor for a Mie void with radius $R = \SI{100}{\nano \meter}$ centered at the origin as a function of the size parameter $kR$ and distance of the emitter $r_d$. We distinguish between the Purcell enhancement for either parallel (a) or perpendicular (b) dipole orientation relative to the void surface. (c) Plot for the Purcell enhancement factor in the case of a freely placed emitter, which is depicted in the small geometry on the left-hand side of the plot. The dipole moment of the emitter is always oriented to the right direction.}
\label{fig: Purcell enhancement}
\end{figure}

We use a multipole expansion series for the electric field to express $F_\parallel $ and $F_\perp $ as a series of multipole coefficients $a_l$ and $b_l$. For a dipole emitter with vacuum wavelength $\lambda$ at distance $r_d$ from the Mie void with radius $R$, the enhancement factors are~\cite{Hohenester2020,10.1063/1.453317}
\begin{align}
	\begin{split}
		F_\perp &=\frac{3}{2} \sum_{l=1}^\infty l(l+1)(2l+1)\left|\frac{j_l(kr_\text{d}) + a_lh_l(kr_\text{d})}{kr_\text{d}} \right|^2,
	\end{split}
\end{align}
and
\begin{align}
	\begin{split}
		F_\parallel &=\frac{3}{4} \sum_{l=1}^\infty (2l+1) \biggl(\left|j_l(kr_\text{d}) + a_lh_l(kr_\text{d}) \right|^2 \\ 
		&+\left| \frac{\psi_l'(kr_\text{d}) + b_l \xi_l' (kr_\text{d}) }{kr_\text{d}}\right|^2   \biggr),
	\end{split}
\end{align}
where again $k$ is the magnitude of the wave vector inside the host medium. 
If $\vb{p}$ is located inside the void, the radiation field outside (in the host) can be computed again via the multipole expansion series method. However it is no longer a linear combination of $\vb{E}_{\text{sca}}$ and $\vb{E}_{\text{dip}}$ \cite{PhysRevA.13.396}.
The resulting enhancement factors are \cite{10.1063/1.453317}
\begin{align}
\begin{aligned}
F_\parallel &= \frac{3}{4}\,\frac{n_\text{host}^3}{n_\text{sphere}}
\sum_{l=1}^{\infty} (2l+1)
\biggl(
\left|
\frac{\psi_l(\eta kr_{\text{d}})'}{\eta kr_{\text{d}} D_l}
\right|^{2}
\\&
+\frac{1}{(n_\text{sphere}n_\text{host})^2}
\frac{j_l^{2}(\eta kr_{\text{d}})}{|D_l'|^{2}} \biggr),
\end{aligned}
\end{align}
and
\begin{align}
	\begin{split}
		F_\perp = \frac{3}{2}\,\frac{n_\text{host}^3}{n_\text{sphere}}
\sum_{l=1}^{\infty}
l(l+1)\,(2l+1)\,
\frac{j_l^{2}(\eta k r_\text{d})}{(\eta k r_\text{d})^{2}\,|D_l|^{2}} ,
	\end{split}
\end{align}
with the abbreviations 
\begin{align}
		D_l &= n^2_\text{sphere}\,j_l(kR)\,\xi_l'(kR)
		- n^2_\text{host}\,h_l(kR)\,\psi_l'(\eta kR),
\end{align}
and
\begin{align}
		D_l'&=  \,j_l(kR)\,\xi'_l(kR)
		-\,h_l(kR)\,j_l'(\eta kR).
\end{align}
Figure~\ref{fig: Purcell enhancement} (a)-(b) show $F_{\parallel}$ and $F_{\perp}$ as a function of $kR$ and the distance $r_d$ from the void. It can be seen that for an emitter inside the void, emission is enhanced up to a factor of 5, where we have cut the upper limit of the colorbar for illustration purposes. Far from the surface, the strongest enhancement is observed at $kR$ values close to those of the Mie void modes, whose spectral locations are marked by the gray dashed lines. This is further confirmed in Fig.~\ref{fig: Purcell enhancement}(c), where we plot $F(\vb{p})$ at different spatial locations inside and outside the void~\cite{Hohenester2020}. 
\nocite{*}
\clearpage
\bibliography{apssamp123}

\end{document}